\documentclass[11pt]{article}
\usepackage[letterpaper,margin=1in]{geometry}
\usepackage[T1]{fontenc}
\usepackage[utf8]{inputenc}
\usepackage{mathptmx}
\usepackage{amsmath,amssymb}
\usepackage{graphicx}
\usepackage{booktabs}
\usepackage{multirow}
\usepackage{url}
\usepackage[hidelinks]{hyperref}
\usepackage{caption}
\usepackage{subcaption}
\usepackage{authblk}
\newcommand{\mbssim}{\mbox{MB-SSIM}}
\newcommand{\cpd}{c/deg}
\DeclareMathOperator*{\Ex}{\mathbf{E}}

\newcommand{\fitFiftyP}{0.8389}   \newcommand{\fitFiftyS}{0.8365}
\newcommand{\fitHundP}{0.8227}    \newcommand{\fitHundS}{0.8091}
\newcommand{\fitPoolP}{0.8343}    \newcommand{\fitPoolS}{0.8323}

\newcommand{\symFiftyP}{0.8367}   \newcommand{\symHundP}{0.8248}
\newcommand{\symPoolP}{0.8330}
\newcommand{\ssFiftyE}{0.9532}  \newcommand{\msFiftyE}{0.8403}  \newcommand{\fitFiftyE}{0.8263}
\newcommand{\ssHundE}{0.8438}   \newcommand{\msHundE}{0.8747}   \newcommand{\fitHundE}{0.8709}
\newcommand{\ssPoolE}{0.9604}   \newcommand{\msPoolE}{0.9180}   \newcommand{\fitPoolE}{0.8579}
\newcommand{\unifFifty}{0.7581} \newcommand{\unifHund}{0.6965}  \newcommand{\unifPool}{0.7165}
\newcommand{\csfFifty}{0.7837}  \newcommand{\csfHund}{0.7889}   \newcommand{\csfPool}{0.7925}

\newcommand{\csfLFifty}{0.8219} \newcommand{\csfLHund}{0.8092}  \newcommand{\csfLPool}{0.8205}

\title{\textbf{\mbssim{}: A Multi-Band Variant of SSIM,\\
       Tunable to Viewing Conditions}}
\author{Yuriy Reznik}
\affil{Massachusetts Institute of Technology\\ \texttt{yreznik@mit.edu}}
\date{}

\begin{document}
\maketitle

\begin{abstract}
\noindent
SSIM, as practically deployed, contains a hidden two-band model:
although it was introduced as a departure from channel-based vision
models, it amounts to a product of one similarity functional applied to
a low-pass and a high-pass version of the two images. This two-band split, however,
is defined in \emph{pixels}, so the range of spatial frequencies that
each band covers, as seen by the viewer, changes with the display and
the viewing distance. Making this model explicit and expressing it in cycles per degree
gives a version of SSIM that takes the viewing conditions into account
while preserving its simplicity. The resulting metric, \mbssim{}, keeps
SSIM's per-patch product of similarity ratios; the bands are placed where psychophysics
places the visual system's channels, and the viewing geometry enters
through a single number, the angular sampling density in pixels per
degree. On the CID:IQ dataset, which rates the same 690 images at two
viewing distances, one set of sub-band weights fitted jointly and evaluated on
held-out content reaches a pooled Pearson correlation (PLCC) of \fitPoolP{} against $0.787$
for SSIM and $0.808$ for MS-SSIM; on LIVE Release~2, calibrating the
geometry yields $0.9564$ against $0.9457$ and $0.9499$. As a
cross-verification, the CID:IQ weights applied unchanged to the
Brightcove--Kingston multi-screen video dataset, which rates the same
clips on a television, a tablet and a phone, give a pooled PLCC of $0.844$
across devices where SSIM and MS-SSIM, which must return one number per
clip, fall to $0.62$ and $0.59$. Accuracy is
flat from three channels to six and insensitive to filter shape, window
and peak placement; of the bank's parameters only the channel bandwidths
matter, and the psychophysically measured bandwidths are a safe choice.
The parameter the result is most sensitive to is SSIM's own stabilising
constant: reducing it to a third of its standard value raises the pooled
PLCC to $0.857$. Source code reproducing the results is available.
\end{abstract}


\section{Introduction}
\label{sec:intro}

SSIM~\cite{wang2004ssim} was presented as a deliberate step away from the
vision models of its day (the channel decompositions, contrast-sensitivity
weightings and error summations of Daly, Lubin and
Watson~\cite{daly1993vdp,lubin1995sarnoff,watson1987cortex}) towards a
single local statistic computable in a few lines of code. The step
succeeded: SSIM and its multi-scale extension MS-SSIM~\cite{wang2003msssim}
are among the most widely deployed full-reference metrics in the imaging
industries, and serve as standard differentiable objectives in
learning-based image processing~\cite{zhao2017losses,ding2021comparison}.
But the departure left out something the vision models had: no
viewing distance, display size or pixel density appears anywhere in SSIM's
definition, even though perceived quality depends on all
three~\cite{westerink1989subjective,catellier2012mobile,barman2022parametric,barman2023multiscreen}.
The same encoded picture viewed on a phone at arm's length and on a
television across a room does not receive the same opinion score, yet SSIM
assigns it one number.

The premise of this paper is that this can be addressed without giving
up SSIM's simplicity, because \emph{SSIM already contains a hidden band
model}.
As shown in~\cite{reznik2023ei}, practically deployed variants of SSIM
closely approximate the product of one and the same similarity functional
applied to a low-pass and a high-pass version of the two images, in
other words a two-band filterbank metric with unit weights. The bands are
hidden inside SSIM's local statistics: the local mean is a pixel of a
low-pass filtered image, and the local variance and covariance are second
moments of the complementary high-pass signal. And because the split is
produced by a Gaussian window of fixed \emph{pixel} width, its crossover
frequency lands wherever the viewing geometry sends it: $3.5$~\cpd{} on an HD television at three picture
heights, $7.1$~\cpd{} on a UHD panel at the same distance, $1.97$ and
$3.95$~\cpd{} at the two viewing distances of the CID:IQ
database~\cite{liu2014cidiq} used below. SSIM thus does perform a
frequency analysis, but in pixel units, so that the result follows the
pixel grid rather than the viewer.

\begin{figure}[!t]
\centering
\includegraphics[width=0.94\textwidth]{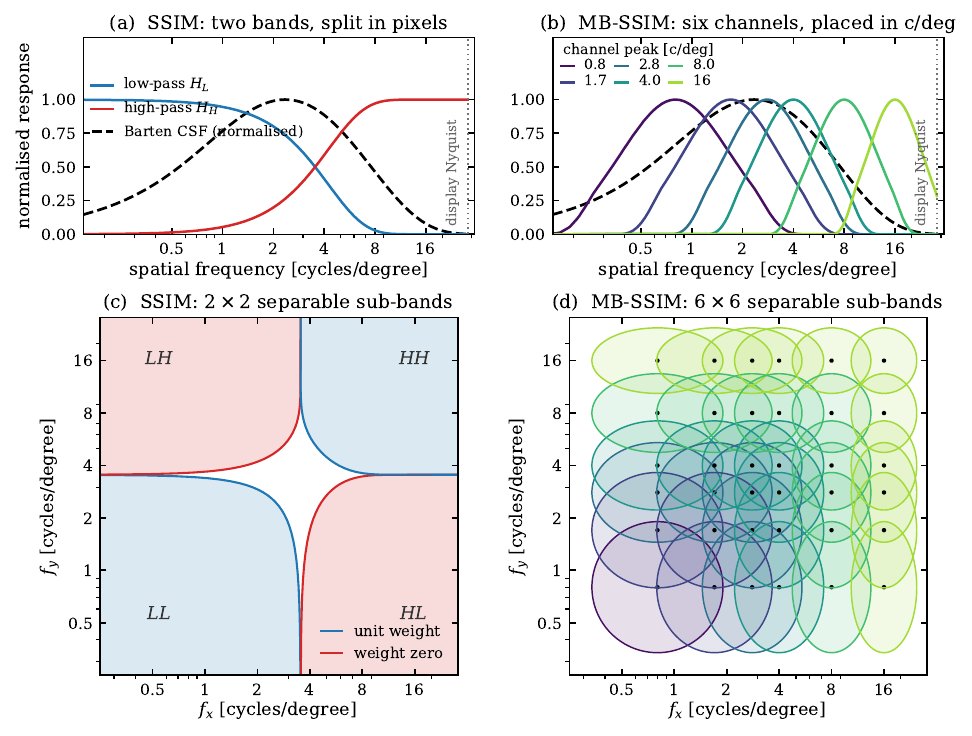}
\caption{The hidden model and the explicit one, at a canonical
high-definition setup ($1920\times1080$ from three picture heights,
$\mathit{ppd}=56.6$ pixels per degree of visual angle). (a)~SSIM's implicit low-pass/high-pass split
($\sigma=3$ pixels) and (b)~\mbssim{}'s six channels at the peaks and
bandwidths of Wilson et al.~\cite{wilson1983oblique}, both against
Barten's contrast sensitivity model~\cite{barten1999csf} (dashed).
(c,~d)~The separable two-dimensional sub-bands they generate, contoured at
half amplitude.}
\label{fig:banks}
\end{figure}

This observation is the starting point of the paper: \emph{make the
hidden band model explicit, and express it in cycles per degree}. Once the bands are explicit, nothing forces there to be
two of them, or forces the crossover to sit at a pixel count; the
psychophysical literature has specified the visual system's
spatial-frequency channels, their number, peaks and bandwidths, for
decades~\cite{wilson1983oblique,watson1981labelled,graham1989analyzers}.
And once they are specified in cycles per degree, the entire viewing
geometry enters through a single number, the angular sampling density
$\mathit{ppd}$, and the same metric is adapted to a phone, a monitor
or a television by changing that one input.
Figure~\ref{fig:banks} shows the two decompositions side by side: SSIM's
hidden pair of bands, and the explicit six-channel bank that replaces it.
The structure of the metric, a per-patch product of similarity ratios
averaged over the image, is untouched. That is the sense in which
\mbssim{} makes SSIM viewing-aware while preserving its practical
simplicity.

That viewing conditions matter has long been recognised; what has been
less available is a way to account for them that is as simple as the
metrics in common use. The classical vision models take the geometry as
an input but are considerably more
complex~\cite{daly1993vdp,lubin1995sarnoff}; Barten's square-root
integral~\cite{barten1990sqri,barman2021sqri} is the closest precedent
for what we do, since it puts the geometry on the frequency axis,
integrating a display's modulation transfer against the contrast
sensitivity function in cycles per degree, but it scores a display chain,
not an image; VMAF maintains separately
trained models for handheld, HD and UHD
viewing~\cite{bampis2026vmafv1}; parametric approaches model quality as a
closed-form function of viewing angle and angular resolution, either as a
correction applied to a viewing-blind metric
score~\cite{barman2022parametric} or as a stand-alone model of quality
against resolution~\cite{barman2022gwr}; in the first case the
correction acts on a number from which the frequency content has already
been removed. \mbssim{} instead puts the geometry where SSIM has been
keeping its band model all along, inside the decomposition itself, at the
cost of one extra input.

The rest of the paper takes the first steps. We realise the explicit
bank with the six channel peaks and bandwidths measured by Wilson et
al.~\cite{wilson1983oblique} (Section~\ref{sec:design}); we argue that one
set of sub-band weights should serve all viewing geometries, because the
angular specification has already absorbed the geometric variation
(Section~\ref{sec:universal}); we validate on CID:IQ~\cite{liu2014cidiq},
the one public image database that rates the same images at two viewing
distances, and on LIVE Release~2~\cite{sheikh2006statistical}, the home
database of both baselines, and cross-verify the CID:IQ weights, applied
unchanged, on the Brightcove--Kingston multi-screen video
dataset~\cite{barman2023multiscreen}, which rates the same clips on a
television, a tablet and a phone (Section~\ref{sec:results}). Source
code for all experiments is available~\cite{nbssimcode}. None of these
choices is claimed to be optimal; the aim is to show that the explicit,
viewing-aware form can be reached at little cost in simplicity.

\section{From the hidden band model to \mbssim{}}
\label{sec:design}

\subsection{The hidden band model}
\label{sec:hidden}

In its general form SSIM is a product of three factors comparing
luminance, contrast and structure,
\begin{equation}
\mathrm{SSIM}(x,y) \;=\; l(x,y)^{\alpha}\, c(x,y)^{\beta}\, s(x,y)^{\gamma},
\label{eq:ssimgen}
\end{equation}
with
\begin{equation*}
l = \frac{2\mu_x\mu_y + C_1}{\mu_x^2+\mu_y^2+C_1},\quad
c = \frac{2\sigma_x\sigma_y + C_2}{\sigma_x^2+\sigma_y^2+C_2},\quad
s = \frac{\sigma_{xy} + C_3}{\sigma_x\sigma_y + C_3},
\end{equation*}
where $\mu$, $\sigma^2$ and $\sigma_{xy}$ are the local mean, variance
and covariance, computed over an $11\times11$ circularly symmetric
Gaussian window of standard deviation $1.5$~pixels, and
$C_1=(K_1L)^2$, $C_2=(K_2L)^2$, $C_3$ are stabilising constants for
dynamic range $L$. The metric as practically deployed (the original
reference implementation, with $K_1=0.01$ and $K_2=0.03$, and every
widely used successor) is the special case $\alpha=\beta=\gamma=1$,
$C_3=C_2/2$, in which the $\sigma_x\sigma_y$ factors cancel between $c$
and $s$ and the product collapses to two factors:
\begin{equation}
\mathrm{SSIM}(x,y) \;=\;
   \frac{2\mu_x\mu_y + C_1}{\mu_x^2+\mu_y^2+C_1}\;\cdot\;
   \frac{2\sigma_{xy} + C_2}{\sigma_x^2+\sigma_y^2+C_2}.
\label{eq:ssim}
\end{equation}
The image score is the average of~(\ref{eq:ssim}) over all positions.

Now define the symmetric similarity functional
\begin{equation}
\xi(a,b) \;=\; \frac{2\,\Ex[ab] + C}{\Ex[a^2] + \Ex[b^2] + C},
\label{eq:xi}
\end{equation}
with $\Ex[\cdot]$ the same Gaussian-weighted local average and $C$ a
stabilising constant of the same form as above. The local mean
$\mu_x$ is a pixel of a low-pass filtered image $x_L$; the residual
$x-\mu_x$ is a pixel of the complementary high-pass image $x_H$; and
$\sigma_x^2$, $\sigma_{xy}$ are second moments of those high-pass signals.
The two factors of~(\ref{eq:ssim}) are therefore the \emph{same}
functional applied to two different bands, and
\begin{equation}
\mathrm{SSIM}(x,y) \;\approx\; \xi(x_L,y_L)\cdot\xi(x_H,y_H),
\label{eq:twoband}
\end{equation}
with a Gaussian split at $\sigma\approx3$ pixels reproducing SSIM to
within about $0.015$ over a range of
distortions~\cite{reznik2023ei}. This is the hidden band model: two bands, unit
weights, crossover fixed in pixels. Two different smoothing operations are
conflated in~(\ref{eq:ssim}) because the same window performs both; the
band split answers \emph{which} frequencies are compared, the local
expectation inside $\xi$ answers \emph{over what neighbourhood}. In what
follows the first is assigned to an explicit filterbank and the second
stays, as in SSIM, inside $\xi$.

\subsection{The explicit metric}
\label{sec:metric}

The explicit bank is built from $N$ one-dimensional band-pass filters
$h_1,\dots,h_N$, the \emph{channels}, with channel $i$ centred at spatial
frequency $f_i$. Filtering the rows of an image with $h_i$ and its
columns with $h_j$,
\begin{equation}
x_{ij} \;=\; h_j \ast_{\mathrm{col}} \big(h_i \ast_{\mathrm{row}} x\big),
\label{eq:subband}
\end{equation}
isolates the \emph{sub-band} $x_{ij}$: the part of the image whose
horizontal detail sits near $f_i$ and whose vertical detail sits near
$f_j$. The $N$ channels thus yield $N^2$ oriented sub-bands, with
sub-band $(i,j)$ responding most strongly to structure at radial
frequency $\sqrt{f_i^2+f_j^2}$ and orientation $\arctan(f_j/f_i)$
(Figure~\ref{fig:banks}c,d). With
$x_{ij}[p]$ denoting the $11\times11$ Gaussian-weighted patch of sub-band
$x_{ij}$ at pixel $p$, exactly as in SSIM, \mbssim{} applies $\xi$ to each
sub-band pair at each location and combines the results as a weighted
product:
\begin{equation}
\mbssim(x,y) \;=\; \frac{1}{|P|}\sum_{p\in P}\;
   \prod_{i=1}^{N}\prod_{j=1}^{N}
   \xi\big(x_{ij}[p],\,y_{ij}[p]\big)^{\,w_{ij}},
\label{eq:mbssim}
\end{equation}
where $P$ is the set of evaluated patch locations. Where the sub-band
indices are clear from context we abbreviate
$\xi(x_{ij}[p],y_{ij}[p])$ to $\xi_{ij}[p]$. The stabilising constant
used in~(\ref{eq:xi}) for sub-band $(i,j)$ is
$C_{ij}=(k_{\max(i,j)}L)^2$, indexed by the finer of the two channels,
with $k_1=K_1$ for the lowest channel and $k_i=K_2$ otherwise, so that
the lowest sub-band retains an SSIM-like $C_1$ and every detail sub-band
a $C_2$. We keep SSIM's values $K_1=0.01$, $K_2=0.03$ throughout, so
that the comparison with SSIM and MS-SSIM is on equal terms;
Section~\ref{sec:ablation} examines them. The $N^2$ numbers $w_{ij}$
are the \emph{sub-band weights}: the exponents that the general
form~(\ref{eq:ssimgen}) always provided and the deployed form froze at
unity. They are the only quantities fitted to
subjective data anywhere in this paper; everything else is determined by
the psychophysics and the viewing geometry. The product in~(\ref{eq:mbssim})
is taken across the sub-bands at each patch location, and only then
averaged over the image, so that the contributions of the sub-bands are
combined locally, where the distortion occurs. The metric is symmetric, bounded by
$1$, and equal to $1$ if and only if $x=y$. For $N=2$, with the two
channels taken as the low-pass and high-pass filters of
Section~\ref{sec:hidden} and weights $w_{11}=w_{22}=1$,
$w_{12}=w_{21}=0$, (\ref{eq:mbssim}) reduces to~(\ref{eq:twoband}),
i.e.\ to SSIM: SSIM is the $N=2$ point of this family.

\begin{figure}[!t]
\centering
\includegraphics[width=0.62\textwidth]{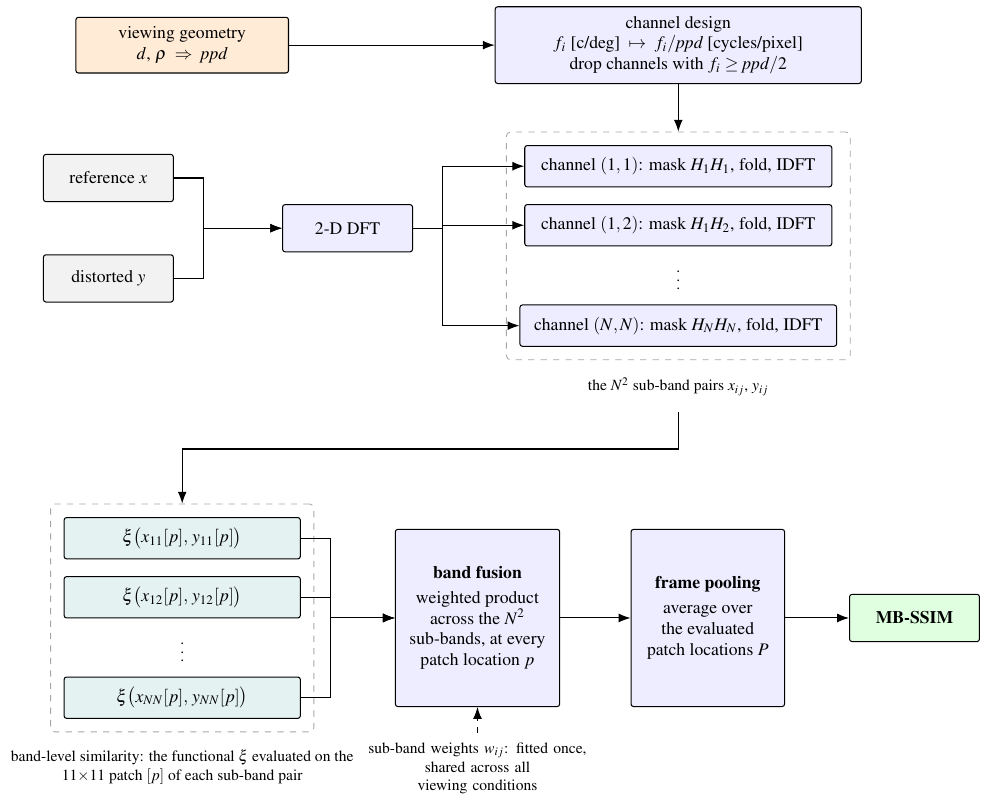}
\caption{Computation of \mbssim{}. The viewing geometry enters only
through $\mathit{ppd}$, which converts channel frequencies from cycles per
degree to cycles per pixel, determines the channel transfer functions,
and decides which channels survive the display Nyquist rate. Each sub-band pair yields a
per-pixel similarity field by~(\ref{eq:xi}); the weighted product
combines them at each location and the final block averages over the
frame.}
\label{fig:flow}
\end{figure}

Figure~\ref{fig:flow} shows the computation. Its structure is SSIM's
(filter, compare locally, average), with the two hidden bands replaced by
an explicit bank.

\subsection{Cycles per degree, and the one number that carries the geometry}
\label{sec:geometry}

\begin{table}[!t]
\caption{Angular sampling density for common viewing setups, and where
SSIM's hidden band split ($\sigma=3$ pixels) falls in cycles per
degree of visual angle. Handheld distances follow measured viewing
distances: about $13$~in for phones~\cite{bababekova2011phones} and
the $18$~in convention of the ITU mobile and tablet
datasets~\cite{barman2022gwr}; television distances are in picture heights
$H$. The field of view is the horizontal angle subtended by the display in
landscape orientation. The split moves by a factor of two across ordinary
conditions; SSIM's weights do not move at all.}
\label{tab:viewing}
\small
\centering
\begin{tabular}{@{}llccrrrr@{}}
\toprule
 & & & & \multicolumn{1}{c}{$\mathit{ppd}$} & \multicolumn{1}{c}{field} & \multicolumn{1}{c}{Nyquist} & \multicolumn{1}{c}{SSIM split} \\
setup & resolution & size & distance & \multicolumn{1}{c}{[px/deg]} & \multicolumn{1}{c}{[deg]} & \multicolumn{1}{c}{[\cpd{}]} & \multicolumn{1}{c}{[\cpd{}]} \\
\midrule
Smartphone & $2532\times1170$ & $6.1''$ & $33$\,cm & $104$ & $24$ & $52.2$ & $6.52$ \\
Tablet & $2360\times1640$ & $10.9''$ & $46$\,cm & $83$ & $28$ & $41.5$ & $5.18$ \\
HDTV & $1920\times1080$ & $55''$ & $3H$ & $57$ & $33$ & $28.3$ & $3.53$ \\
UHD TV & $3840\times2160$ & $55''$ & $1.5H$ & $57$ & $61$ & $28.3$ & $3.53$ \\
UHD TV & $3840\times2160$ & $55''$ & $3H$ & $113$ & $33$ & $56.5$ & $7.06$ \\
\bottomrule
\end{tabular}
\end{table}

\begin{figure}[!t]
\centering
\includegraphics[width=0.84\textwidth]{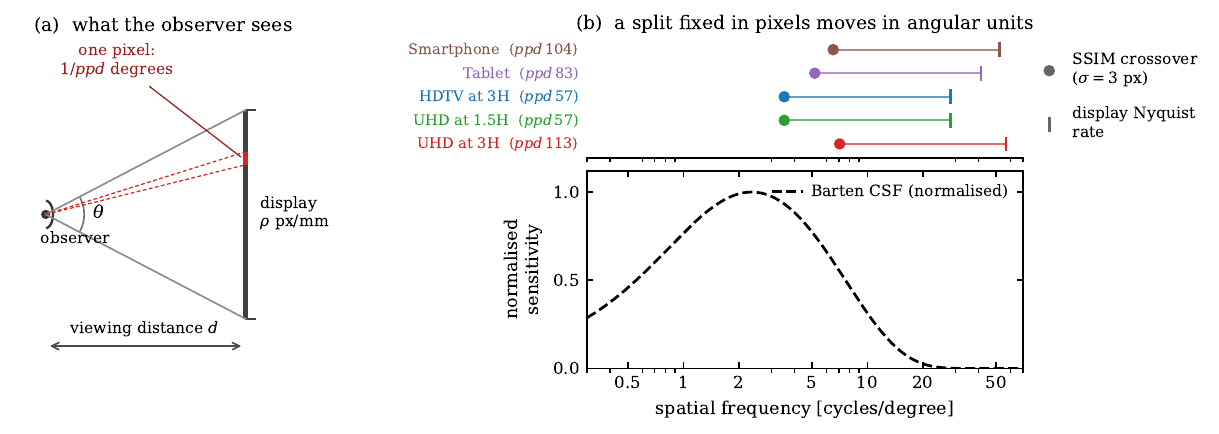}
\caption{(a)~The viewing geometry: distance $d$ and pixel density $\rho$
fix the angular sampling density $\mathit{ppd}$ of~(\ref{eq:ppdintro}).
(b)~Barten's contrast sensitivity function lives on the angular axis and
does not move with the setup; SSIM's pixel-referred band split does.
Each bar runs from where SSIM's $\sigma=3$ pixel crossover lands (dot)
to the display Nyquist rate (tick), for the five setups of
Table~\ref{tab:viewing}.}
\label{fig:geom}
\end{figure}

What the observer sees is set by the viewing distance $d$ and the display
pixel density $\rho$, which together fix the angular sampling density
\begin{equation}
\mathit{ppd} \;=\; \Big[2\arctan\!\big(1/(2 d \rho)\big)\Big]^{-1}
\label{eq:ppdintro}
\end{equation}
in pixels per degree of visual angle~\cite{barman2022parametric}
(Figure~\ref{fig:geom}a). An image feature at $\nu$ cycles per pixel
reaches the eye at $\nu\cdot\mathit{ppd}$ cycles per degree, and nothing
above the display Nyquist rate $\mathit{ppd}/2$ reaches it at all.
Everything the metric needs to know about the setup enters through this
one number.

Table~\ref{tab:viewing} makes the problem quantitative and
Figure~\ref{fig:geom}(b) makes it visible: across five ordinary setups
SSIM's hidden crossover wanders from $3.5$ to $7.1$~\cpd{} while the
weights either side of it stay where they were fitted. A bank specified in
cycles per degree inverts the situation. A channel at $f$~\cpd{} sits at
$f/\mathit{ppd}$ cycles per pixel, so the decomposition is recomputed from
$\mathit{ppd}$ for every setup, and a channel whose peak lies at or above
$\mathit{ppd}/2$ is dropped. This is a modelling statement rather than
a numerical convenience: a distortion the display cannot reproduce cannot
be seen.

\subsection{Realising the bank}
\label{sec:bank}

For the explicit bank we take the most direct reading of the
psychophysics: $N=6$ channels with peaks at
$0.8, 1.7, 2.8, 4, 8, 16$~\cpd{} and half-amplitude bandwidths of
$2.5, 2.2, 1.9, 1.7, 1.5, 1.25$ octaves, the foveal mechanisms measured by
Wilson, McFarlane and Phillips~\cite{wilson1983oblique} and confirmed,
in count and bandwidth, by other psychophysical methods and by
physiology~\cite{watson1981labelled,devalois1982tuning}.

Each channel is a log-Gabor filter~\cite{field1987loggabor},
\begin{equation}
H_i(f) \;=\; \exp\!\left(-\frac{\log_2^2(f/f_i)}{2s_i^2}\right)\,
             T\!\left(\log_2(f/f_i);\,B_i\right),
\qquad s_i=\frac{b_i}{2\sqrt{2\ln 2}},
\label{eq:loggabor}
\end{equation}
with $b_i$ the bandwidth in octaves. The factor $T$ is a raised-cosine
taper: it is unity out to $0.6B_i$ octaves from the centre and falls to
zero at $B_i$ octaves, with $B_i=b_i$, giving each channel compact
support in frequency; the realised peaks stay within $1\%$ of the
targets and the bandwidths and peak gains within $10^{-4}$. The
decomposition is carried out in
the discrete Fourier transform (DFT) domain: the filter transfer functions
are sampled directly onto the transform grid, so the realised bank is the
requested bank, and each sub-band is inverse-transformed at its own
critical sampling rate before $\xi$ is formed. At
the $50$\,cm distance of CID:IQ~\cite{liu2014cidiq} ($\mathit{ppd}=31.6$,
Nyquist $15.8$~\cpd{})
the $16$~\cpd{} channel is dropped and five channels remain; at $100$\,cm
all six survive.

\subsection{One set of weights for every viewing geometry}
\label{sec:universal}

The angular specification has a consequence that is central to the design:
the sub-band weights should \emph{not} depend on the viewing geometry.
Whatever the setup, the $4$~\cpd{} channel is looking at detail that the
viewer sees at $4$~\cpd{}, so the sub-band it produces means the same
thing to the eye in every condition, and there is no reason for its
weight to change. This is also how the visual system is usually
modelled: the channels themselves do not change when the observer moves;
what changes is which spatial frequencies the display delivers to them,
and that is what the $\mathit{ppd}$ mapping and the Nyquist criterion
capture. (What the filterbank cannot absorb, namely display luminance,
ambient illumination and picture size, would have to enter through the
weights. Barten's model~\cite{barten1999csf} already describes two of
these: its contrast sensitivity depends on luminance, and on picture size
through the number of grating cycles the eye can integrate, a term that
is active for pictures below about $12^\circ$ (the regime of Westerink
and Roufs' size experiments~\cite{westerink1989subjective}) and is worth under
$8\%$ on the channels that matter between the $12.6^\circ$ and
$24.9^\circ$ fields of CID:IQ's two distances, which is one more reason
a single weight set can serve both. The databases used here hold luminance fixed,
so the extension is not needed for the present experiments.)

We therefore fit a \emph{single} set of weights jointly across viewing
conditions and require it to serve all of them. This is also the
stronger experimental design, since fitting each condition separately
could not distinguish the filterbank tracking the geometry from the fit
absorbing the difference. Conditions may retain different channel counts
(five at $50$\,cm, six at $100$\,cm), but the surviving channels are the
same channels at the same cycles per degree, so each condition indexes
into one master $6\times6$ weight matrix and a dropped channel
contributes no factor to the product. Since
$\mbssim = \frac{1}{|P|}\sum_p \exp\big(\sum_{i,j} w_{ij} \log\xi_{ij}[p]\big)$
with the $\log\xi_{ij}$ independent of $w$, the $\log\xi_{ij}[p]$ values
are computed once per database and cached, and every candidate weight
vector then costs only a few dot products; optimisation is by the
Nelder--Mead simplex method with basin hopping.

\section{Experimental validation}
\label{sec:results}

\subsection{Databases and protocol}
\label{sec:protocol}

\begin{figure}[!t]
\centering
\includegraphics[width=0.88\textwidth]{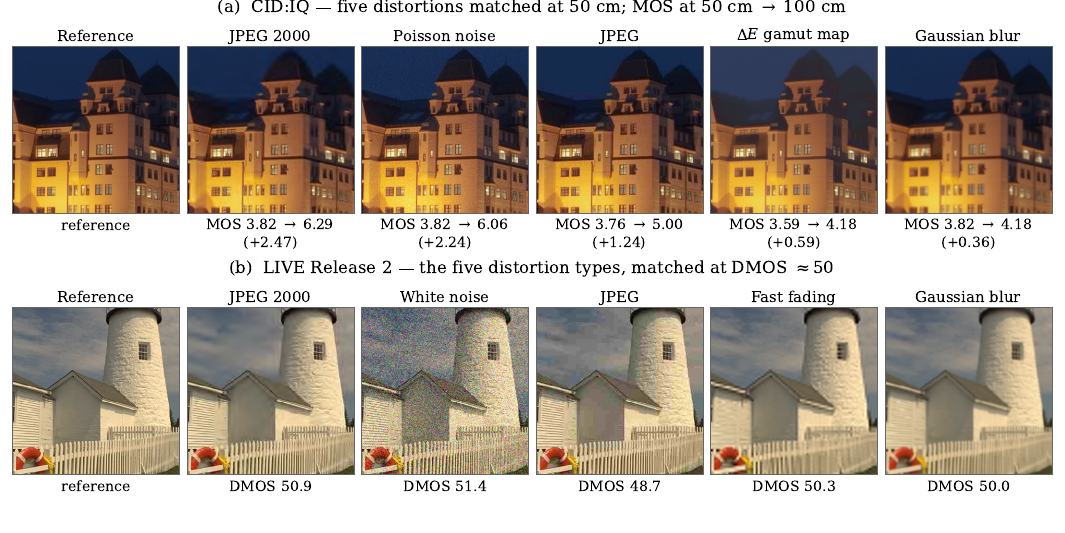}
\caption{Example content, cropped to $320\times320$ pixels at native
resolution.
(a)~CID:IQ: one reference and five distortions rated nearly equal at
$50$\,cm and spanning $2.11$ MOS at $100$\,cm; the reordering follows the
spectrum of each distortion. (b)~LIVE Release~2: one reference and the
five distortion types at matched subjective severity.}
\label{fig:examples}
\end{figure}

\textbf{CID:IQ}~\cite{liu2014cidiq} contains 23 reference images and 690
distorted versions (six distortion types at five levels) rated by 17
observers \emph{twice}: at $50$ and $100$\,cm. It is, to our knowledge,
the only public full-reference \emph{image} database that repeats the same
stimuli at two viewing distances (for video, the multi-screen set
of~\cite{barman2023multiscreen} rates the same content on a television, a
tablet and a phone), and hence the most direct test available of the
claim made here. Its own evaluation script fixes the display at $92$~dpi,
giving $\mathit{ppd}=31.612$ and $63.223$ at the two distances; we use
these values unchanged, so no part of the geometry mapping is fitted. The
distance effect in the scores is large and frequency-dependent, as
expected: the mean opinion scores (MOS) rise on average by
$0.606$ from $50$ to $100$\,cm, and Figure~\ref{fig:examples}(a) shows
five distortions of one reference rated within $0.23$ MOS of each other at
$50$\,cm and across a $2.11$-point range at $100$\,cm: high-frequency
distortions (JPEG2000 ringing, Poisson noise) recover more than two points when the observer
steps back and their energy falls below the threshold of visibility,
low-frequency ones (blur, gamut mapping) recover a third of a point. A metric with no
viewing-geometry input returns one number per image and therefore one
ordering for both distances.

\textbf{LIVE Release~2}~\cite{sheikh2006statistical} is the database on
which SSIM and MS-SSIM were originally validated: 29 reference images, 779
distorted versions across five distortion types, scored as difference mean
opinion scores (DMOS) on a $1$--$100$ scale
(Figure~\ref{fig:examples}b). Its viewing conditions were held constant
across all its experiments but are not published, so we treat
$\mathit{ppd}$ as the one unknown of the setup and calibrate it
(Section~\ref{sec:live}).

\textbf{Protocol.} A metric score $s$ is mapped to the subjective scale
with the VQEG five-parameter logistic~\cite{vqeg2000}
\begin{equation}
Q(s) \;=\; \beta_1\!\left(\tfrac12 -
   \frac{1}{1+e^{\beta_2 (s-\beta_3)}}\right) + \beta_4 s + \beta_5,
\label{eq:logistic}
\end{equation}
fitted by nonlinear least squares; the Pearson correlation (PLCC) and the
root-mean-square error (RMSE) are computed after the fit, so that a metric
is not penalised for a monotone non-linearity, and the Spearman rank
correlation (SRCC) is computed on the raw scores, which is the procedure
of the SSIM and MS-SSIM
papers~\cite{wang2004ssim,wang2003msssim,sheikh2006statistical}. All three
metrics are evaluated on the luma channel (BT.601 coefficients) over the
whole frame. SSIM uses Wang's published settings including automatic
pre-downsampling; MS-SSIM uses the published five-scale weights. Because \mbssim{} has fitted
parameters and the baselines do not, every \mbssim{} number below is
\emph{held out}: five-fold cross-validation with folds split by reference
image, weights refitted from scratch within each fold jointly across both
viewing distances, and evaluated on the held-out references only. Our
SSIM and MS-SSIM implementations were checked against Wang's MATLAB code
(agreement to $5\times10^{-13}$); the code for all experiments is
available~\cite{nbssimcode}. The computational cost is modest: the DFT
implementation scores an $800\times800$ pair at $\mathit{ppd}=63.2$ in
$1.3$\,s on one core of a 2.8\,GHz Xeon.

\subsection{Results on CID:IQ}
\label{sec:cidiq}

\begin{table}[!t]
\centering
\caption{Results on CID:IQ. \mbssim{} numbers are held out: five-fold
cross-validation, folds split by reference image, one set of sub-band
weights fitted jointly across both viewing distances. Best in each row
in bold; RMSE on the MOS scale after the logistic fit.}
\label{tab:main}
\small
\begin{tabular}{@{}lccc@{}}
\toprule
& SSIM & MS-SSIM & \mbssim{} \\
\midrule
\multicolumn{4}{@{}l}{\emph{(a) PLCC}}\\
50\,cm  & 0.7783 & 0.8331 & \textbf{\fitFiftyP} \\
100\,cm & \textbf{0.8328} & 0.8201 & \fitHundP \\
pooled  & 0.7867 & 0.8081 & \textbf{\fitPoolP} \\
\addlinespace[3pt]
\multicolumn{4}{@{}l}{\emph{(b) SRCC}}\\
50\,cm  & 0.7790 & 0.8339 & \textbf{\fitFiftyS} \\
100\,cm & \textbf{0.8191} & 0.8135 & \fitHundS \\
pooled  & 0.7822 & 0.8060 & \textbf{\fitPoolS} \\
\addlinespace[3pt]
\multicolumn{4}{@{}l}{\emph{(c) RMSE (lower is better)}}\\
50\,cm  & \ssFiftyE & \msFiftyE & \textbf{\fitFiftyE} \\
100\,cm & \textbf{\ssHundE} & \msHundE & \fitHundE \\
pooled  & \ssPoolE & \msPoolE & \textbf{\fitPoolE} \\
\bottomrule
\end{tabular}
\end{table}

\begin{figure}[!t]
\centering
\includegraphics[width=0.88\textwidth]{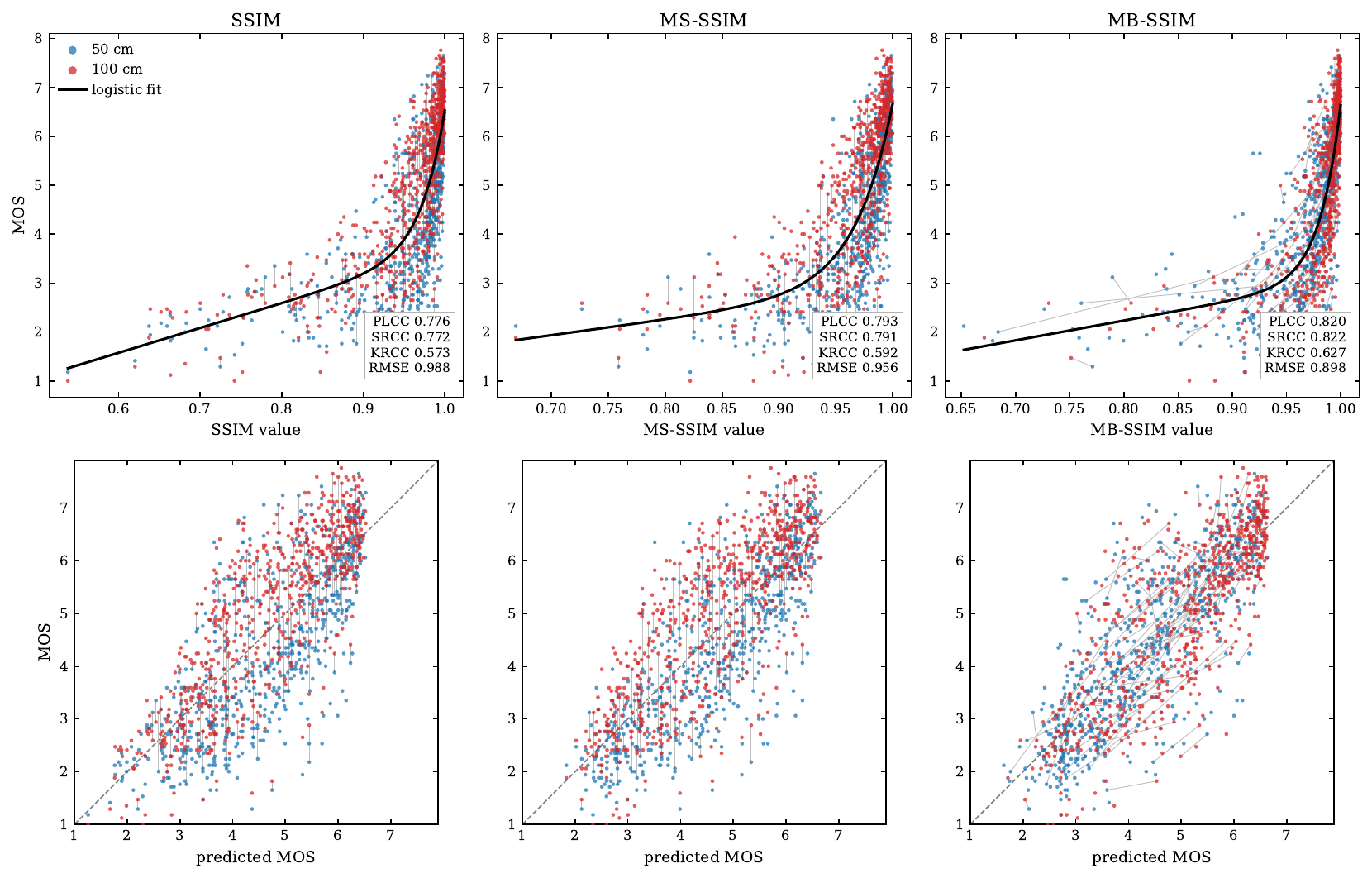}
\caption{Goodness of fit on CID:IQ, all 1380 ratings, one logistic fit
per column. Top: MOS against the raw metric value; bottom: MOS against
the predicted MOS. Thin segments join the two ratings of the same image:
vertical for the viewing-blind metrics, slanted for \mbssim{}.}
\label{fig:fit}
\end{figure}

Table~\ref{tab:main} gives the results, and they follow the expected
pattern. At each distance separately, \mbssim{}
matches the better of the two baselines: it is ahead of SSIM by $0.061$
PLCC and level with MS-SSIM at $50$\,cm, and within about $0.01$ of both
at $100$\,cm.
More telling is what happens \emph{between} the distances: SSIM and
MS-SSIM trade first and last place, because each carries weights tuned to
one implicit geometry, while \mbssim{} stays at or near the top in both
conditions with a single parameter vector. And pooled across both
distances, the situation a multi-device system faces, where one score
must mean one quality, the difference is larger:
\mbssim{} reaches PLCC \fitPoolP{} against $0.787$ for SSIM and $0.808$
for MS-SSIM, reducing the residual RMSE by $11\%$ and $6.5\%$. A metric
without a viewing-geometry input is at a disadvantage in this pooled test
by construction, since it returns one value against two different
targets.
Figure~\ref{fig:fit} shows the mechanism directly: the segments joining
the two ratings of the same image are exactly vertical for SSIM and
MS-SSIM, whose score for an image is the same at both distances, and
slanted for \mbssim{}, whose filterbank is re-derived from
$\mathit{ppd}$.

\subsection{Results on LIVE Release 2}
\label{sec:live}

We sweep $\mathit{ppd}$, refitting the weights at each value under the
same five-fold protocol. Accuracy peaks at
$\mathit{ppd}=80$, and Table~\ref{tab:livebest} reports that operating
point: PLCC $0.9564$ against $0.9457$ for SSIM and $0.9499$ for MS-SSIM,
with RMSE falling from $8.85$ and $8.51$ to $7.96$ DMOS. The margin is
smaller than on CID:IQ, as one would expect: LIVE is a single-condition
database, the case the baselines' fixed weights were chosen for, and the
explicit bank is still slightly ahead. Weights fitted on CID:IQ and
transferred to LIVE without refitting reach only parity with SSIM; with
$23$ reference images identifying $36$ weights, cross-database transfer
would need more training data than is available here.

\begin{table}[!t]
\centering
\caption{LIVE Release 2 at the calibrated geometry ($\mathit{ppd}=80$),
weights refitted on LIVE, held-out five-fold cross-validation with folds
split by reference image.}
\label{tab:livebest}
\small
\begin{tabular}{@{}lccc@{}}
\toprule
 & SSIM & MS-SSIM & \mbssim{} \\
\midrule
PLCC & 0.9457 & 0.9499 & \textbf{0.9564} \\
SRCC & 0.9469 & 0.9496 & \textbf{0.9556} \\
RMSE & 8.85 & 8.51 & \textbf{7.96} \\
\bottomrule
\end{tabular}
\end{table}

\subsection{Cross-verification: the Brightcove--Kingston multi-screen dataset}
\label{sec:multiscreen}

The weights of Table~\ref{tab:main} were fitted on CID:IQ. As a
cross-verification on a different medium, with different devices,
observers and laboratory, and with no parameter refitted, we applied
them to the Brightcove--Kingston (BC--KU) multi-screen
dataset~\cite{barman2023multiscreen}. CID:IQ varies one thing, the
distance; this dataset varies the device: the same $27$ HEVC
renditions of three $1080$p sources ($540$p, $720$p and $1080$p at three
quality levels) were rated by the same $26$ observers on a $55''$ UHD
television at $3H$, a $12.4''$ tablet at $18$~in and a $6.1''$ phone at
$12$--$14$~in, giving $81$ mean opinion scores. The device effect is large
and systematic: the mean MOS is $2.85$ on the television, $3.52$ on the
tablet and $4.17$ on the phone, and the phone is rated above the television
for all $27$ clips. On the $1080$p grid at which the renditions are
compared, the geometry of~(\ref{eq:ppdintro}) gives effective angular
sampling densities of $56.5$ (television, video scaled $2\times$), $57.4$
(tablet, scaled $1.46\times$) and $97.9$~\cpd{} (phone, shown $1{:}1$). The
television and the tablet are therefore almost the same angular
presentation, and the phone is a different one.

We applied all three metrics frame-wise, as SSIM and MS-SSIM routinely are
for video, on every twentieth frame of each rendition upscaled bicubically
to $1080$p against its source, and averaged over frames. \mbssim{} was
evaluated at each device's effective $\mathit{ppd}$ with the sub-band
weights of Table~\ref{tab:main}, fitted on CID:IQ \emph{still images} and
\emph{applied without any fitting on this data}. Table~\ref{tab:msds}
reports PLCC after a logistic fit per device, which lets every metric
absorb the device preference, and after one logistic over all $81$
points, which does not.

\begin{table}[!t]
\centering
\caption{Cross-verification on the Brightcove--Kingston multi-screen
dataset~\cite{barman2023multiscreen}: PLCC at the clip level, $27$ clips
per device. Left, a separate logistic fit per device;
right, one logistic over all $81$ (clip, device) points, and over the two
device pairs. \mbssim{} uses the CID:IQ weights of Table~\ref{tab:main}
unchanged: nothing in this table was fitted on this dataset.}
\label{tab:msds}
\small
\begin{tabular}{@{}lcccc|ccc@{}}
\toprule
 & \multicolumn{4}{c|}{logistic per device} & \multicolumn{3}{c}{one logistic} \\
 & phone & tablet & TV & mean & all 81 & TV+phone & TV+tablet \\
\midrule
SSIM                                  & 0.896 & 0.880 & 0.859 & 0.878 & 0.617 & 0.543 & 0.768 \\
MS-SSIM                               & 0.848 & 0.837 & 0.838 & 0.841 & 0.592 & 0.518 & 0.741 \\
\mbssim{}, uniform weights            & 0.785 & 0.788 & 0.826 & 0.800 & 0.730 & 0.750 & 0.709 \\
\mbssim{}, CID:IQ weights             & \textbf{0.933} & \textbf{0.884} & \textbf{0.870} & \textbf{0.895} & \textbf{0.844} & \textbf{0.887} & \textbf{0.786} \\
\bottomrule
\end{tabular}
\end{table}

Three observations follow. First, the weights transfer: with no
fitting at all, the weights learned from still images at two distances
carry over to video across three devices; per device, \mbssim{} is
ahead of SSIM on all three and ahead of MS-SSIM by $0.05$ on average.
Second, in the case the metric is intended for, when one score must mean
one quality on every device, the metrics without a geometry input lose
most of their correlation: SSIM drops from $0.878$ to $0.617$,
because it returns the same number for a clip whose MOS differs by $1.3$
between the television and the phone, while \mbssim{} keeps $0.844$ and
orders the phone above the television for all $27$ clips, as the observers
did. On the two geometrically distinct devices, television and phone, the
gap is $0.887$ against $0.543$. Third, the limit of a geometric model is
equally visible: on the television and the tablet, whose angular
presentations differ by less than two percent, every metric scores
between $0.71$ and $0.79$, because the $0.7$-point MOS gap between those two devices is not
geometric (it is the preference for the hand-held device that the
dataset's authors report) and only a per-device mapping can absorb it.
That weights fitted on $23$ still images carry over to video on three
devices suggests that they encode something about the visual system
rather than the particulars of one database.

\subsection{Ablations: what carries the result, and how small the metric can be}
\label{sec:ablation}

\begin{table}[!t]
\caption{Channel-count sweep on CID:IQ (held-out, one weight set jointly
across both distances; PLCC). All banks span $0.8$--$16$~\cpd{}: the
first four place their peaks log-uniformly (Figure~\ref{fig:sweep}), the
last is the Wilson bank of Section~\ref{sec:bank}. `channels' counts
those below the display Nyquist rate at each distance.}
\label{tab:sweep}
\small
\centering
\begin{tabular}{@{}lccccc@{}}
\toprule
bank & weights & channels, $50/100$\,cm & 50\,cm & 100\,cm & pooled \\
\midrule
$N=3$ & 9 & 2\,/\,3 & 0.8401 & 0.8252 & 0.8384 \\
$N=4$ & 16 & 3\,/\,4 & 0.8384 & 0.8280 & 0.8382 \\
$N=5$ & 25 & 4\,/\,5 & 0.8392 & 0.8264 & 0.8367 \\
$N=6$ & 36 & 5\,/\,6 & 0.8385 & 0.8229 & 0.8345 \\
$N=6$, Wilson peaks & 36 & 5\,/\,6 & 0.8389 & 0.8227 & 0.8343 \\
\bottomrule
\end{tabular}
\end{table}

\paragraph{The weights carry the fit; the angular bank carries the
geometry.} Table~\ref{tab:ablation} compares five ways of setting the
sub-band weights $w_{ij}$ in~(\ref{eq:mbssim}), all on the six-channel
bank and all evaluated on the same held-out folds. In the first row
nothing is fitted: every sub-band receives the same weight,
$w_{ij}=2/N^2$, chosen so that the weights sum to $2$ just as SSIM's two
unit exponents do. In the second row nothing is fitted either, but the
weights follow the psychophysics: $w_{ij}$ is proportional to Barten's
contrast sensitivity~\cite{barten1999csf} $S(f_{ij})$ at the sub-band's
radial frequency $f_{ij}=\sqrt{f_i^2+f_j^2}$, evaluated at the luminance
of the CID:IQ display, and again normalised to sum to $2$. The third row adds one
fitted number to that: the weight of the lowest sub-band, $w_{11}$. In the
last row all $36$ weights are free parameters, fitted to the subjective
scores, and the row before it fits them under the symmetry constraint
$w_{ij}=w_{ji}$, which treats horizontal and vertical detail alike and
leaves $21$ free values.

The table reads as a decomposition of what the fit finds. Uniform
weights reach a pooled PLCC of \unifPool{}; the contrast sensitivity
function alone, with nothing fitted, \csfPool{}; the contrast sensitivity
function plus one number, \csfLPool{}; all $36$ weights free, \fitPoolP{}
(and the symmetric fit gives up almost nothing, \symPoolP{}). The fitted
weights are therefore, to within $0.014$ PLCC, Barten's contrast
sensitivity plus a term SSIM has always had: band-pass channels transmit
nothing of the image mean, and the weight on the lowest sub-band takes
over the role of SSIM's luminance comparison. This is the intended
division of labour: the bank, specified in cycles per degree, carries the
viewing geometry; the weights carry the perceptual emphasis, the role
the exponents of~(\ref{eq:ssimgen}) were always meant to play, and that
emphasis is close to what the psychophysics predicts.

\begin{table}[!t]
\centering
\caption{Five parameterisations of the sub-band weights on CID:IQ
(six-channel Wilson bank; PLCC on the same held-out folds as
Table~\ref{tab:main}). The second column counts the parameters fitted to
subjective scores. CSF: weights proportional to Barten's contrast
sensitivity at the sub-band frequency.}
\label{tab:ablation}
\small
\begin{tabular}{@{}lcccc@{}}
\toprule
sub-band weights & fitted & 50\,cm & 100\,cm & pooled \\
\midrule
uniform, $w_{ij}=2/N^2$            &  0 & \unifFifty & \unifHund & \unifPool \\
CSF, $w_{ij}\propto S(f_{ij})$     &  0 & \csfFifty  & \csfHund  & \csfPool \\
CSF $+$ lowest sub-band $w_{11}$   &  1 & \csfLFifty & \csfLHund & \csfLPool \\
symmetric, $w_{ij}=w_{ji}$         & 21 & \symFiftyP & \symHundP & \symPoolP \\
all weights free                   & 36 & \fitFiftyP & \fitHundP & \fitPoolP \\
\bottomrule
\end{tabular}
\end{table}

\begin{figure}[!t]
\centering
\includegraphics[width=0.68\textwidth]{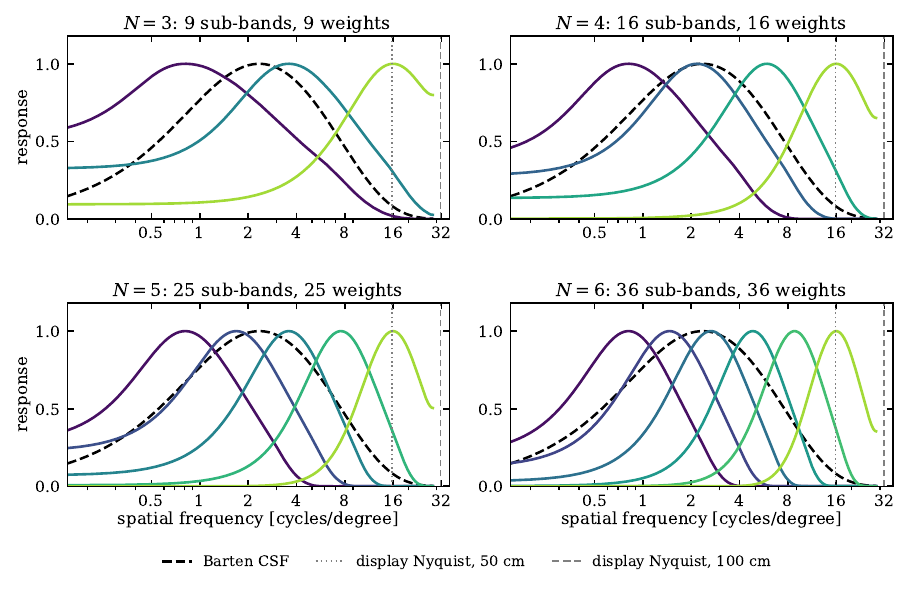}
\caption{The channel-count sweep: all banks span the same range, with
peaks log-uniform between $0.8$ and $16$~\cpd{} and bandwidths scaled by
$6/N$. Vertical lines mark the display Nyquist rate at the two CID:IQ
distances; the $16$~\cpd{} channel is dropped at $50$\,cm in every bank.}
\label{fig:sweep}
\end{figure}

\paragraph{Three channels do the work of six.} Six channels is the design
point because that is what the psychophysics reports, but the count should
not be read too literally: masking and summation experiments cannot
distinguish six discrete channels from a dense continuum of similar
filters, of which six is the parsimonious discrete
sampling~\cite{graham1989analyzers}. How much, then, does the count
matter here? Sweeping it over banks spanning the same
$0.8$--$16$~\cpd{} range
(Table~\ref{tab:sweep}, Figure~\ref{fig:sweep}) moves held-out pooled
PLCC by less than the fold-to-fold spread of $0.039$: $0.8384$, $0.8382$, $0.8367$,
$0.8345$ for $N=3,4,5,6$, across a fourfold range in the number of
fitted weights. A three-channel bank built by merging Wilson's channels in
pairs, whose top peak at $11.3$~\cpd{} survives the Nyquist rate at
both distances, gives the best figure of the sweep, $0.8407$, with nine
weights. The cheapest configuration is thus also among the best; the
practical form of the metric is small, and the six-channel Wilson bank
should be read as the reference design rather than as a cost the approach
requires. (With only two
viewing geometries in the data there is little for a finer bank to do; a
database spanning several could order the banks differently.)

\paragraph{Bandwidth matters; placement, shape and window matter little.} The count
is not the only design choice, so we also perturbed the six-channel bank
itself (Table~\ref{tab:perturb}; $28$ configurations in all, every one
under the protocol above). Three choices are free: the filter
shape (log-Gabor with or without the taper, or a difference of Gaussians),
the statistics window ($\sigma$ from $1.0$ to $2.5$) and linear versus
logarithmic spacing all move held-out pooled PLCC by $0.003$ or less, none
of it significant. Placement is nearly free and asymmetric: jittering every
peak independently by up to a third of an octave never helps and costs at
most $0.004$; sliding the whole bank down costs accuracy consistently across
folds ($-0.006$ at $-\tfrac13$ octave, $p<0.01$), sliding it up is neutral
to marginally positive. Bandwidth is the one bank parameter the metric is
sensitive to, and the dependence is one-sided: narrowing the channels to
$60\%$ of Wilson's values loses $0.024$, the largest effect of any bank
ablation in this paper, while widening them by $25\%$ gains a non-significant
$0.005$ and wider still gains nothing. A bank with a constant $1.5$-octave
bandwidth, which leaves Wilson's high channels unchanged and narrows
only the low ones from $2.5$ and $2.2$ octaves, loses $0.015$, and in the
jointly drawn banks the sign of the outcome is set by the bandwidth scale
alone: every draw wider than Wilson's bank matches or beats the reference,
every draw narrower loses significantly. The wide low-frequency mechanisms
that distinguish Wilson's measurements from a generic octave bank thus
matter, and the psychophysically measured bandwidths lie on the safe
side of this dependence.

\begin{table}[!t]
\centering
\caption{Perturbing the six-channel bank itself, on CID:IQ (held-out
five-fold cross-validation, joint fit across both distances; pooled PLCC).
All configurations, including the reference, use a reduced search budget
(one restart, two basin-hopping rounds), which costs the reference $0.003$
relative to Table~\ref{tab:main}; comparisons within the table are
like-for-like. $\Delta$ is the fold-paired difference from the reference
for the best and worst configuration of each family.}
\label{tab:perturb}
\small
\begin{tabular}{@{}lcccccc@{}}
\toprule
perturbation & configs & min & mean & max & worst $\Delta$ & best $\Delta$ \\
\midrule
Wilson bank (reference)                       & 1 & \multicolumn{3}{c}{0.8312} & --- & --- \\
all peaks shifted $\pm\tfrac16,\pm\tfrac13$ octave & 4 & 0.8248 & 0.8305 & 0.8360 & $-0.006^{**}$ & $+0.005$ \\
each peak jittered, $U(\pm\tfrac13)$ octave    & 6 & 0.8276 & 0.8294 & 0.8310 & $-0.004^{**}$ & $-0.000$ \\
peaks spaced linearly in frequency            & 1 & \multicolumn{3}{c}{0.8293} & $-0.002$ & \\
all bandwidths scaled $\times0.6$--$1.5$      & 4 & 0.8073 & 0.8251 & 0.8359 & $-0.024^{*}$ & $+0.005$ \\
constant $1.5$-octave bandwidth               & 1 & \multicolumn{3}{c}{0.8167} & $-0.015$ & \\
filter shape (no taper, taper $0.8$, DoG)     & 3 & 0.8278 & 0.8304 & 0.8327 & $-0.003$ & $+0.002$ \\
statistics window $\sigma=1.0,\,2.5$          & 2 & 0.8305 & 0.8316 & 0.8327 & $-0.001$ & $+0.002$ \\
peaks and bandwidths drawn jointly            & 6 & 0.8157 & 0.8261 & 0.8329 & $-0.016^{*}$ & $+0.002$ \\
\bottomrule
\multicolumn{7}{@{}l}{\footnotesize $^{*}p<0.05$, $^{**}p<0.01$, paired over the five folds.}
\end{tabular}
\end{table}

\paragraph{The stabilising constants act as thresholds, and are the
most sensitive parameter.} Every result so far keeps SSIM's $K_1=0.01$
and $K_2=0.03$. In~(\ref{eq:xi}) the constant acts as a threshold: a
sub-band whose energies fall below it contributes nothing, and one well
above it is compared relative to its own energy, i.e.\ to the picture's
own contrast in that band. Expressed as a Weber contrast (the
sub-band's rms amplitude relative to the local mean luminance), SSIM's
$K_2$ puts that threshold near $10\%$ at mid-grey, far above the
detection thresholds of the visual system. Scaling both constants and refitting
(Table~\ref{tab:constants}) has a larger effect than any bank
parameter: a third of SSIM's values raises the held-out pooled PLCC
from $0.831$ to $0.857$, three times them lowers it to $0.782$, and the
optimum lies near $K_2\approx0.01$, a threshold of $3$--$4\%$ contrast.
At that level almost nothing needs fitting: with the weights set to
Barten's contrast sensitivity as in Table~\ref{tab:ablation} and only
the lowest sub-band's weight fitted, the metric reaches $0.856$,
matching the $36$-weight fit, and chosen on the training folds alone
(which select $K_2=0.01$ in every fold), so the gain survives
cross-validation. We have not re-based the paper on the lower constants,
since the comparisons above are fairest with SSIM's own values and the
optimum was found on one database; but the constants are the second
hidden parameter of SSIM, and the natural place for the contrast
sensitivity function's thresholds, and hence for display luminance, to
enter.

\begin{table}[!t]
\centering
\caption{Scaling SSIM's stabilising constants $(K_1,K_2)=(0.01,0.03)$
on CID:IQ (six-channel Wilson bank; held-out pooled PLCC, same folds
and budget as Table~\ref{tab:perturb}). Second column: weights
proportional to Barten's contrast sensitivity with the lowest sub-band's
weight chosen on the training folds, as in Table~\ref{tab:ablation}.
Third column: all $36$ weights refitted.}
\label{tab:constants}
\small
\begin{tabular}{@{}lccc@{}}
\toprule
constants & $K_2$ & CSF $+$ lowest sub-band (1 fitted) & all weights free (36) \\
\midrule
$\times 1/10$ & 0.003 & 0.811 & 0.824 \\
$\times 1/5$ & 0.006 & 0.847 & --- \\
$\times 1/3$ & 0.010 & \textbf{0.856} & \textbf{0.857} \\
$\times 1/2$ & 0.015 & 0.852 & 0.855 \\
$\times1$ (SSIM's values, used throughout) & 0.030 & 0.821 & 0.831 \\
$\times3$           & 0.090 & 0.762 & 0.782 \\
\bottomrule
\end{tabular}
\end{table}

\begin{figure}[!t]
\centering
\includegraphics[width=0.68\textwidth]{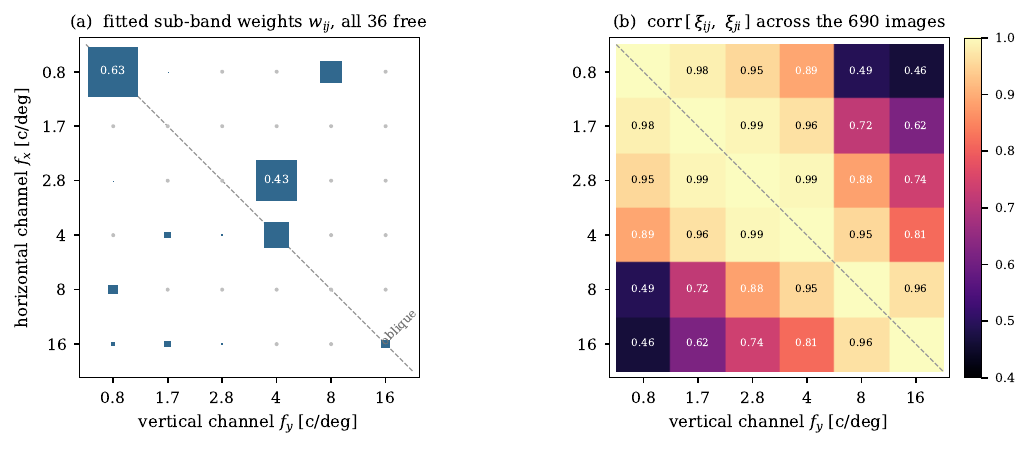}
\caption{(a)~The weight matrix fitted jointly across both viewing
distances, all 36 entries free; square area proportional to $w_{ij}$.
(b)~Correlation between per-image mean $\log\xi_{ij}$ and its mirror
$\log\xi_{ji}$: the largest asymmetries in (a) occur where $r>0.99$ in
(b), so they carry no information.}
\label{fig:weights}
\end{figure}

\paragraph{What the fitted weights mean.} The fitted matrix
(Figure~\ref{fig:weights}a) concentrates its mass on the lowest sub-band
and near $4$~\cpd{}, where contrast sensitivity peaks, the two
ingredients that Table~\ref{tab:ablation} showed to account for it: the
lowest sub-band alone predicts MOS at a pooled PLCC of $0.70$, the
contrast-sensitivity weights without it at $0.77$, the two together at
$0.82$, the same complementarity as SSIM's luminance and structure terms.
The matrix's apparent asymmetry should not be over-read: mirror sub-bands
$(i,j)$ and $(j,i)$ are nearly identical predictors on natural images
(correlation above $0.99$ for near-diagonal pairs,
Figure~\ref{fig:weights}b), so which of the two the search loads is not
determined by the data. This is why the symmetry constraint of
Table~\ref{tab:ablation} costs only about $0.001$ PLCC, and why we regard
the symmetric form as the parsimonious choice.

\section{Conclusions}
\label{sec:conclusions}

SSIM, as practically deployed, is a two-band metric: it applies one
similarity functional to a low-pass and a high-pass version of the two
images and multiplies the results. Because the split is defined in
pixels, the spatial frequencies the two bands cover, as seen by the
viewer, change with every display and viewing distance. \mbssim{} makes
this band model explicit and places the bands in cycles per degree,
where psychophysics places the channels of the visual system. SSIM's
per-patch product of similarity ratios is kept, and the viewing geometry
enters through a single input, the angular sampling density in pixels
per degree.

The metric was tested on three databases. On CID:IQ, which rates 690
images at two viewing distances, one set of sub-band weights fitted
jointly and evaluated on held-out content reaches a pooled PLCC of
\fitPoolP{}, against $0.787$ for SSIM and $0.808$ for MS-SSIM. On LIVE
Release~2, calibrating the viewing geometry gives $0.956$ against
$0.946$ and $0.950$. On the Brightcove--Kingston multi-screen video
dataset, which rates the same clips on a television, a tablet and a
phone, the CID:IQ weights applied without refitting are ahead of SSIM
and MS-SSIM on every device, and when one score must hold across all
three devices they give a pooled PLCC of $0.844$ where SSIM and MS-SSIM,
which return one number per clip whatever the device, fall to $0.62$
and $0.59$.

The metric is small and tolerant of design choices. Three channels and
nine weights match a six-channel bank; accuracy is insensitive to filter
shape, window and peak placement, and of the bank's parameters only the
channel bandwidths matter, and the psychophysical values are a safe
choice. The one parameter to which the result is sensitive
is inherited from SSIM: reducing its stabilising constants to a third of
their standard values raises the pooled PLCC on CID:IQ to $0.857$, and
at that setting the sub-band weights can be set directly from the
contrast sensitivity function with a single fitted number. Those
constants are also the natural place for display luminance to enter the
metric, through the thresholds that Barten's model supplies at every
frequency and luminance, and this is the extension we plan next.

Source code reproducing the results of this paper is
available~\cite{nbssimcode}. The band model was always part of SSIM;
once it is made explicit, SSIM's simplicity and the knowledge of spatial
vision accumulated in psychophysics need not be alternatives.


\end{document}